\documentclass[12pt,a4paper]{article}

\usepackage[english]{babel}
\usepackage[utf8]{inputenc}
\usepackage{graphicx}
\usepackage{fullpage}
\usepackage{color}
\usepackage[table]{xcolor}
\usepackage{latexsym,amsmath,verbatim}
\usepackage{amssymb}
\usepackage{amsfonts}
\usepackage{array}
\usepackage[utf8]{inputenc}
\usepackage{rotating}
\usepackage{makeidx}
\usepackage{tabularx}
\usepackage{savesym}
\savesymbol{iint}
\usepackage[affil-it]{authblk}

\title{Coupling news sentiment with web browsing data improves prediction of intra-day price dynamics}

\author[1,**]{ Gabriele Ranco}
\author[2]{Ilaria Bordino}
\author[3,4]{Giacomo Bormetti}
\author[1,5,6]{Guido Caldarelli}
\author[3,4]{Fabrizio Lillo}
\author[4,7,*]{Michele Treccani}

\affil[1]{IMT Institute for Advanced Studies, Piazza San Francesco 19, 55100 Lucca, Italy}
\affil[2]{Yahoo Labs, Barcelona, Spain}
\affil[3]{Scuola Normale Superiore, Piazza dei Cavalieri 7, 56126 Pisa, Italy}
\affil[4]{QUANTLab, Via Pietrasantina 123, 56122 Pisa, Italy}
\affil[5]{ISC-CNR, Via dei Taurini 19, 00185 Roma, Italy}
\affil[6]{London Institute for Mathematical Science, South St. 35 Mayfair, London W1K 2XF, UK}
\affil[7]{Mediobanca S.p.A, Piazzetta E. Cuccia 1, 20121 Milano, Italy}
\vspace{0.5cm}

\affil[*]{\small{ The opinions expressed here are solely those of the authors and do not represent in any way those of their employers.}} 
\affil[**]{\small{Correspondence should be addressed to G.R. (\emph{gabriele.ranco@gmail.com})}}

\vspace{0.5cm}

\begin{document}
\maketitle

\begin{abstract}

  The new digital revolution of big data is deeply changing our capability of understanding society and forecasting the outcome of many social and economic systems. Unfortunately, information can be very heterogeneous in the importance, relevance, and surprise it conveys, affecting severely the predictive power of semantic and statistical methods. Here we show that the aggregation of web users' behavior can be elicited to overcome this problem in a hard to predict complex system, namely the financial market. {Specifically, our in-sample analysis shows that the combined use of sentiment analysis of news and browsing activity of users of Yahoo! Finance greatly helps forecasting intra-day and daily price changes of a set of 100 highly capitalized US stocks traded in the period 2012-2013}. Sentiment analysis or browsing activity when taken alone have very small or no predictive power. {Conversely, when considering a {\it news signal} where in a given time interval we compute the average sentiment of the clicked news, weighted by the number of clicks,  we show that for nearly 50\% of the companies such signal Granger-causes hourly price returns}.  Our result indicates a ``wisdom-of-the-crowd'' effect that allows to exploit users' activity to identify and weigh properly the relevant and surprising news, enhancing considerably the forecasting power of the news sentiment.

\end{abstract}

%\keywords{financial markets | complex systems | data science | computational social science}

\section{Introduction}
The recent technological revolution with widespread presence of computers, users and media connected by Internet has created an unprecedented situation of data deluge, changing dramatically the way in which we look at social and economic sciences.  As people increasingly use the Internet for information such as business or political news, online activity has become a mirror of the collective consciousness, reflecting the interests, concerns, and intentions of the global population with respect to various economic, political, and cultural phenomena. Humans' interactions with technological systems are generating massive datasets documenting collective behaviour in a previously unimaginable fashion \cite{king2011ensuring,vespignani2009predicting}. By properly dealing with such data collections, for instance representing them by means of network structures \cite{bonanno2004networks,caldarelli2007scale}, it is possible to extract relevant information about the evolution of the systems considered (i.e. trading \cite{tumminello2005tool}, disease spreading \cite{achrekar2011predicting,tizzoni2012real}, political elections  \cite{caldarelli2014multilevel}).

A particularly interesting case of study is that of the financial markets. Markets can be seen as collective decision making systems, where exogenous (news) as well as endogenous (price movements) signals convey valuable information on the value of a company. Investors continuously monitor these signals in the attempt of forecasting future price movements.  Because of their trading based on these signals, the information is incorporated into prices, as postulated by the Efficient Market Hypothesis  \cite{malkiel1970efficient}. Therefore the flow of news and data on the activity of investors can be used to forecast price movements. The literature on the relation between news and price movement is quite old and vast. In order to correlate news and price returns one needs to assess whether the former is conveying positive or negative information about a company, a particular sector or on the whole market. This is typically done with the sentiment analysis, often performed with dedicated semantic algorithms as described and reviewed in the Methods Section.

In this paper, we combine the information coming from the sentiment conveyed by public news with the browsing activity of the users of a finance specialized portal  to forecast price returns at daily and intra-day time scale. To this aim we leverage a unique dataset consisting of a  fragment of the log of Yahoo! Finance, containing the news articles displayed on the web site and the respective number of ``clicks'', i.e. the visualizations made by the users. Our analysis considers 100 highly capitalized US stocks in a one-year period between 2012 and 2013.

For each of these companies we build a signed time series of the sentiment expressed in the related news. The sentiment expressed in each article mentioning a company is weighted {by the number of views of the article}. In our dataset each click action is associated with a timestamp recording the exact point in time when such action took place. Thus we are able to construct time series at the time resolution of the minute. To the best of our knowledge, this is the first time that an analysis like the one described in this paper is conducted at such intra-day granularity. The main idea behind this approach is that the sentiment analysis gives information on the news, while the browsing volume enable us to properly weigh news according to the attention received from the users.

We find that news on the same company are extremely heterogeneous in the number of clicks they receive, an indication of the huge difference in their importance and the interest these news generate on users. For 70\% of the companies examined, there is a significant correlation between the browsing volumes of financial news related to the company, and its traded volumes or absolute price returns. More important, we show that for more than 50\% of the companies (at hourly time scale), and for almost 40\% (at daily time scale), the click weighted average sentiment time series Granger-cause price returns, indicating a rather large degree of predictability.

\section*{Data}  \label{sec:methods}
\subsection*{Stocks considered}
Our analysis is conducted on highly capitalized stocks belonging to the Russell 3000 Index traded in the US equity markets, which we monitor for a period of one year between 2012 and 2013. Among all companies, we selected the 100 stocks with the largest number of news published onYahoo! Finance during the investigated period.
The ticker list of the investigated stocks with a distinctive numerical company identifier follows: 1~KBH, 2~LEN, 3~COST,	4~DTV,	5~AMGN,	6~YUM, 7~UPS, 8~V, 9~AET, 10~GRPN, 11~ZNGA, 12~ABT,	13~LUV,	14~RTN,	15~HAL,	16~ATVI, 17~MRK,	18~GPS,	19~GILD, 20~LCC,	21~NKE,	22~MCD,	23~UNH,	24~DOW,	25~M, 26~CBS,	27~COP,	28~CHK,	29~CAT,	30~HON,	31~TWX,	32~AIG,	33~UAL,	34~TXN,	35~BIIB, 36~WAG,	37~PEP,	38~VMW,	39~KO,	40~QCOM, 41~ACN,	42~NOC,	43~DISH, 44~BBY,	45~HD,	46~PG, 47~JNJ,	48~AXP,	49~MAR,	50~TWC,	51~UTX,	52~MA,	53~BLK,	54~EBAY, 55~DAL,	56~NWSA, 57~MSCI, 58~LNKD, 59~TSLA, 60~CVX,	61~AA, 62~NYX,	63~JCP,	64~CMCSA, 65~NDAQ, 66~IT, 67~YHOO, 68~DIS,	69~SBUX, 70~PFE,	71~ORCL, 72~HPQ,	73~S, 74~LMT,	75~XOM,	76~IBM, 77~NFLX, 78~INTC, 79~CSCO, 80~GE, 81~WFC, 82~WMT, 83~AMZN, 84~VOD, 85~DELL, 86~F, 87~TRI, 88~GM, 89~FRT, 90~VZ, 91~FB, 92~BAC, 93~MS, 94~JPM, 95~C, 96~BA, 97~GS, 98~MSFT, 99~GOOG, 100~AAPL.
The numerical identifiers are assigned according to the increasing order of the total number  of published news in Yahoo! Finance.

We considered three main sources of data for the selected stocks:
\subsection*{Market data}
The first source contains information on price returns and trading volume of the stock at the resolution of the minute. We consider different time scales of investigation, corresponding to $1$, $10$, $30$, $65$, and $130$ minutes. The above values are chosen because they are sub-multiple of the trading day in the US markets (from 9:30 AM to 4:00 PM, corresponding to $390$ minutes). For each time scale and each stock we extract the following time series:
\begin{itemize}
  \item $V$, the traded volume in that interval of time,
  \item $R$, the logarithmic price return in the time scale,
  \item $\sigma$, the return absolute value, a simple proxy for the stock volatility.
\end{itemize}
The precise definition of these variables is given in Materials Section. Since trade volumes and absolute price returns are known to display a strong intra-day pattern, we de-seasonalize the corresponding time series (in the same section we provide the details about this procedure). This procedure is necessary in order to avoid the detection of spurious correlation and Granger causality due to the presence of a predictable intra-day pattern.

\subsection*{News data}
The second source of data consists of the news published on Yahoo! Finance together with the time series of the aggregated clicks made by the users browsing each page.
Yahoo! Finance is a web portal for news and data related to financial companies, offering news and information around stock quotes, stock exchange rates, corporate press releases, financial reports, and message boards for discussion.
Providing consumers with a broad range of comprehensive online financial services and information, Yahoo! Finance has consistently been a leader in its category: In May 2008\footnote{\texttt{http://www.comscore.com/Insights/Press-Releases/2008/07/Yahoo!-Finance-Top-Financial} \texttt{-News-and-Research-Site-in-US}} it was the top financial website with 18.5 million U.S. visitors, followed by  AOL Money \& Finance with 15.2 million visitors (up 48 percent) and MSN Money with 13.7 million visitors (up 13 percent). As of today, recent estimates released in July 2015\footnote{\texttt{http://www.niemanlab.org/2015/07/newsonomics-how-much-is-the-financial-times-worth-and} \texttt{-who-might-buy-it/}} confirm that Yahoo! Finance, with more than 72 million visitors, is still the leader finance website in the US, and the fourth in the whole world.

We analyze a portion of the log of Yahoo! Finance, containing news articles displayed on the portal. The articles are tagged with the specific companies (e.g., Google, Yahoo!, Apple, Microsoft) or financial entities (e.g., market indexes, commodities, derivatives) that are mentioned in its text. The dataset analyzed in this work does not consist of public data. It was extracted from a browsing log of the Yahoo! Finance web portal. The log stores all the actions made by the users who visit the website, such as views, clicks and comments on every page displayed on the portal. Specifically, we extracted the news articles displayed on Yahoo! Finance and the respective number of ``clicks'', i.e. the visualizations made by the users. We considered 100 US stocks in a one-year period between 2012 and 2013.

For each considered company we build a signed time series of the sentiment expressed in the related news. The sentiment expressed in each article mentioning a company is weighted {by the number of views of the article}. In our dataset each click action is associated with a timestamp recording the exact point in time when such action took place. Thus we are able to construct time series at the time resolution of the minute. While building the dataset, we observed the corporate policy of Yahoo with respect to the confidentiality of the data and the tools used in this research. Any sensitive identifier of Yahoo user was discarded after the extraction and aggregation process. Moreover our dataset does not store single actions or users, but only aggregated browsing volumes of financial articles displayed on Yahoo! Finance. Although the original log of Yahoo! Finance is proprietary and cannot obviously be shared, for repeatability of our analysis we can provide the browsing-volume time series extracted for the 100 companies as supplementary material.

In order to automatically detect whether the article is conveying positive or negative news on the company, we perform a sentiment analysis. To obtain a sentiment score, we classify each article with {\em SentiStrength} \cite{thelwall2010sentiment}, a state-of-the-art tool for extracting positive and negative sentiment from informal texts. The tool is based on a dictionary of ``sentiment'' words, which are manually picked by expert editors and annotated with a number indicating the amount of positivity or negativity expressed by them. The original dictionary of {\em SentiStrength} is not tailored to any specific knowledge or application domain, thus it is not the most proper choice to compute a \emph{financial} sentiment. To solve this issue, following a practice that is common in most research on sentiment analysis and price returns~\cite{wang2013financial}, we adapt the original dictionary by incorporating a list of sentiment keywords of special interest and significance for the financial domain \cite{loughran2011liability}. In Materials Section we discuss the robustness of this choice as well as the way news are associated to stocks.

Supported by previous research that studied stock price reaction to news headlines~\cite{chan2003stock,tetlock2007giving,mao2011predicting,lillo2012how, martinez2012semantic, reis2015breaking}, we simplify our data processing pipeline by performing the sentiment analysis computation on the title of each article, instead of using its whole content. The main reason for this choice is that the tone of the news is typically highlighted in the title, while the use in the text of many neutral words can increases the noise and reduces the ability of assessing the sentiment. Finally, the choice also depended on the availability of data: the log at our disposal did not always contained the text of the news and this would have forced us to use a significant subsample.

The sentiment score is a simple sign $(-1,0,+1)$ for each news depending on whether there are more positive or negative words in the title.

\subsection*{Browsing Data}
Finally, in our analysis we use the information on the browsing volume, that is, the time series of ``clicks'' that the web users made on each article displayed on Yahoo! Finance {to view its content}. Given that the users' activity on this domain-specific portal proved to provide a clean signal of interest in financial stocks \cite{bordino2014stock}, we exploit it in this work to weight the sentiment of each article on a given financial company. Specifically, we use the number of clicks of an article as a proxy for the level of attention that users gave to that news. By aggregating over a time window the clicks on all the articles, even published earlier, that mention a particular company, it is possible to derive an estimation of the attention around that company.

In summary, for each time scale and for each stock, the variables we extract from the  database are (see Materials Section):
\begin{itemize}
  \item $C$,  the time series of the total number of clicks in a time window,
  \item $S$, the sum of the sentiment of all news related to each company,
  \item $WS$, the sum of the sentiment of all news weighted by the number of clicks.
\end{itemize}
The first quantity $C$ is non negative and measures the level of attention in a given time interval for news about a specific company. The $S$ variable is the usual sentiment indicator employed in numerous studies and provides the aggregated sentiment of the company specific news published in a given time interval. The most important and novel quantity is $WS$, which combines the two previous ones by assigning a sign to each click depending on the sentiment of the clicked news. As for the market variables, we remove the intra-day pattern from the click time series. In fact, both the publication of news \cite{lillo2012how} and the clicking activity of  users \cite{bordino2014stock} show a strong intra-day seasonality. These patterns are probably related to the way humans carry out their activities during the day (e.g. {small activity during lunchtime}, more hectic activity at the beginning or at the end of the business day).

\section*{Methods}
\subsection*{Sentiment Analysis}
Regarding the analysis of search-engine queries, some recent works \cite{bank2011google, bordino2012web, preis2010complex, kristoufek2013can, vlastakis2012information,zhangetal2013em} have studied the relation between the daily number of queries related to a particular stock and the trading volume over the same stock.

An incomplete list of contributions on the role of news includes studies investigating
(i) the relation between exogenous news and price movements \cite{chan2003stock,curme2014quantifying,cutler1998moves, vega2006stock},
(ii) the correlation between high or low pessimism of media and high market trading volume \cite{tetlock2007giving};
(iii) the relation between the sentiment of news, earnings and return predictability \cite{tetlock2008more, schumaker2009textual},
(iv) the role of news in the trading action \cite{lillo2012how,engelberg2012shorts, alanyali2013quantifying, zhang2014internet};
(v) the role of macroeconomic news in the performance of stock returns \cite{birz2011effect},
and (vi) the  high-frequency market reaction to news \cite{gross2011machines}.

For example, in a recent paper \cite{bordino2012web}, related to ours, authors show that daily trading volumes of stocks traded at NASDAQ can be forecasted with the daily volumes of queries related to the same stocks. In another paper a similar analysis shows that an increase in queries predicts higher stock prices in the next two weeks \cite{da2011search}. In \cite{zhangetal2013em} authors test the explanatory power of investor attention -- measured as the search frequency at daily level of stock names in Baidu Index -- for abnormal daily returns and find evidence that the former Granger causes stock price returns in excess with respect to the market index return, whereas there is little evidence for the opposite causal relation. As for social networks and micro-blogging platforms \cite{de2008can}, Twitter data is becoming an increasingly popular choice for financial forecasting. For example some have investigated whether the daily number of tweets predicts  SP 500 stock prices \cite{mao2012correlating, ruiz2012correlating}. A textual analysis approach to Twitter data can be found in other works \cite{mao2011predicting,bollen2011twitter, bollen2011modeling, zhang2011predicting, zheludev2014can} where the authors find clear relations between mood indicators and Dow Jones Industrial Average. Some other authors have used news, Wikipedia data or search trends to predict market movements \cite{curme2014quantifying, preis2012quantifying, preis2013quantifying, moat2013quantifying}.

There are two main critical aspects in the kind of analyses described above. {First, the universe of all the search engine or social network users is probably too large and the fraction of users truly interested in finance is likely quite low. This is particularly true at intraday frequency, investigated in this paper. Second, as we will empirically show below, the universe of news considered is  very heterogeneous in terms of their relevance as a signal of future price movement. For example, in a day there might be several positive but almost irrelevant news and only one negative but very important news on a company. Without weighting the relevance of the news, one could easily draw a wrong conclusion}. The intuition behind the current work is that the number of times a news is viewed by users is a measure of its importance as well as of the surprise it conveys. Moreover the users we consider are not generic, but are those who use one of the most important news and search portals for financial information, namely Yahoo! Finance.

\subsection*{Spearman Correlation}
To overcome these limitations we collected for each stock and for each time scale a total of six time series, namely $V$, $R$, $\sigma$, $C$, $S$, and $WS$, and we study their dependence by making use of two tools. {First, given two time series $X_t$ and $Y_t$, we consider the Spearman's correlation coefficient
  \begin{equation}
    \rho(X,Y)=\frac{\langle {r_X}_t {r_Y}_t\rangle-\langle {r_X}_t\rangle \langle {r_Y}_t\rangle}{\sqrt{(\langle {r_X}_t^2\rangle-\langle {r_X}_t\rangle^2)(\langle {r_Y}_t^2\rangle -\langle {r_Y}_t\rangle^2)}}
  \end{equation}
where ${r_X}_t$ and ${r_Y}_t$ correspond to the rank of the $t$-th realization of the $X$ and $Y$ random variables, respectively, and $\langle \cdot \rangle$ is the time average value. The correlation $\rho(X,Y)$ quantifies the linear contemporaneous dependence without relying on the Normal assumption for $X$ and $Y$}. In order to assess the statistical significance of the measured value we perform a statistical test of the null hypothesis that the correlation is zero by randomizing the time series.

\subsection*{Granger Causality}
Our main goal is testing for the presence of statistical causality between the variables. To this end our second tool is the Granger causality test \cite{granger69}. Granger's test is a common test used in time series analysis to determine if a time series $X_t$ is useful in forecasting another time series $Y_t$.
$X_t$ is said to \emph{Granger-cause} $Y_t$ if $Y_t$ can be better predicted using both the histories of $X_t$ and $Y_t$ rather than using only the history of $Y_t$. The Granger causality can be assessed by regressing $Y_t$ on its own time-lagged values and on those of $X_t$. An F-test is then used to examine whether the null hypothesis that $Y_t$ is not Granger-caused by $X_t$ can be rejected with a given confidence level (in this paper we use a p-value of 5\%).

\section*{Results} \label{sec_results}
The most important aspect of our analysis is to test whether one can forecast financial variables, and more specifically price returns, by using the information on the browsing activity of the users. Namely if, by weighting the sentiment of the clicked news by the number of clicks each news receives, one can improve significantly the predictability of returns.

\subsection*{Heterogenous attention}
\begin{figure}[t]
  \centering
  \includegraphics[width=0.75\textwidth]{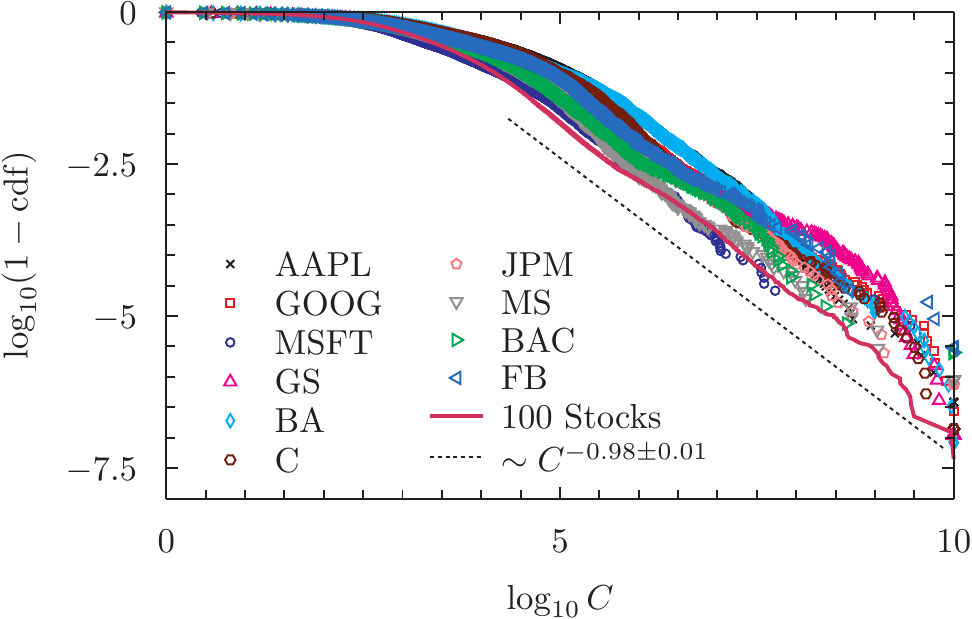}
  \caption{Complementary of the cumulative distribution function of the number of clicks a news receives for the ten assets with the largest number of news and the aggregate portfolio of 100 stocks. Both coordinates have been rescaled by a common factor preserving the power law scaling of the right tail and normalizing the maximum number of clicks to the value $10^{10}$. The dotted line corresponds to a power law with tail exponent fitted from the portfolio time series. We provide details about the standard error and the complete list of tail exponents for all the companies in Materials Section.} \label{heterogeneity}
\end{figure}

The first observation is the extreme heterogeneity of the attention that users of Yahoo! Finance show towards the financial news of a given company. Figure \ref{heterogeneity} shows the complementary of the cumulative distribution function of the number of clicks per news concerning a given stock. There we show the curves for each of the top 10 stocks and for the aggregate of 100 stocks. In all cases the tail of the distribution is very well fit by a power law behavior~\cite{zipf49} with a tail exponent very close to $1$ (see Materials Section) In fact, the mean exponent across all stocks is $1.15\pm 0.30$ and restricting on the top $10$ it is $0.99\pm 0.08$. This indicates that there is a huge heterogeneity in the number of clicks a news receives and therefore in the importance users give to it. It is also a warning that not weighting properly the importance of the news can lead to overstate the importance of the many irrelevant news and to understate the importance of the few really important ones.

\subsection*{Synchronous correlation}
\begin{figure}[t]
  \centering
  \includegraphics[width=0.75\textwidth]{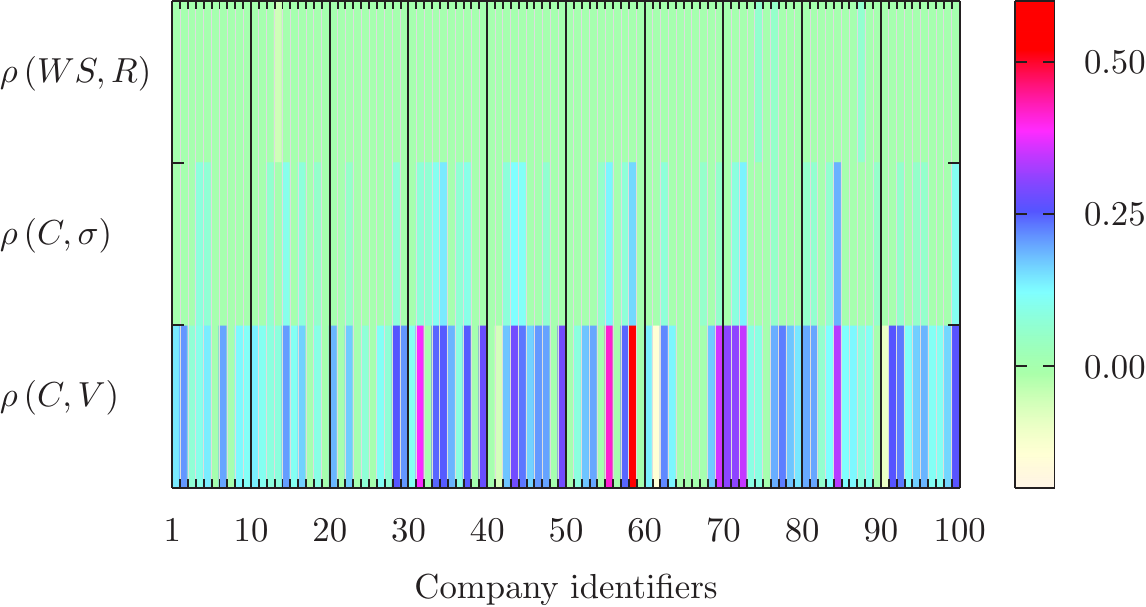}
  \caption{Spearman correlation coefficients for the de-seasonalized time series of all the 100 companies at hourly scale. The x axis reports the list of companies identified by a unique number, as detailed in the main text. Among the several possibilities, we consider only three couples and the color scale corresponds to the level of correlation. We plot those values for which we reject the null of zero correlation at 5\% significance level and equalize non significant values to zero (light green color).} \label{fcor1}
\end{figure}

In order to understand  how the relation between financial and news variables depends on the time scale, we perform a  synchronous correlation analysis. For each of the 100 companies, we compute the Spearman's correlation coefficient $\rho$ between the three sensible pairs made by one ``news'' time series and one ``financial'' time series. Figure \ref{fcor1} summarizes the results for the 65 min time series. The x axis lists the companies, uniquely identified by a number that provides the rank of the company in the order from the least to the most cited one (as measured by the absolute number of associated news). Thus, 1 corresponds to the company KBH with the least number of news, while 100 to the most cited AAPL. We label the y axis with the pairs $(C,V)$, $(C,\sigma)$, and $(WS,R)$, while the color scale indicates the level of correlation. We compute the correlation sampling the original time series every 65 minutes, equalizing to zero those values whose significance does not reject the null hypothesis of zero correlation with 5\% confidence. 
{Figure~\ref{fcor1} shows in general a positive and significant correlation between browsing activity and price volatility and volume, whereas the evidence of linear dependence between sentiment time series and price returns is mild, similarly to the result obtained by Mao \cite{mao2011predicting}. In the fourth row of Table~\ref{allcor1} we report the percentage of the 100 companies for which we reject the null hypothesis of zero correlation at 5\% confidence level and. Since we use multiple correlation tests in order to establish whether their is a significant relationship between key news and online quantities and key market variables, in the Materials Section we report results corrected for multiple hypotheses testing. Applying the conservative correction proposed by Bonferroni, the evidence of linear dependence enterily survives for click, volatility, and trade volume time series, whereas the hypothesis of zero-Spearman's correlation between price returns and weighted sentiment is no more rejected.
\begin{table}
  \caption{Percentage of companies for which we reject the null hypothesis of zero Spearman correlation at 5\% confidence level.}
  \begin{tabular*}{\columnwidth}{@{\extracolsep{\fill}}lrrr}
    Time interval (minutes) &  $\rho(WS,R)$ &  $\rho(C,\sigma)$ &  $\rho(C,V)$ \cr
    \hline
    1        &      7 &       86 &      95 \cr
    10       &      3 &       72 &      90 \cr
    30       &      5 &       54 &      85 \cr
    65       &      4 &       36 &      79 \cr
    130      &      4 &       26 &      76 \cr
    \hline
  \end{tabular*}
  \label{allcor1}
\end{table}
}

\subsection*{Time scale}
In order to investigate how the correlation changes with the time scale, in Table~\ref{allcor1} we also show the percentage of rejection for the 1, 10, 30, and 130 minutes time scales. As a general comment, we observe that the number of companies with a significant correlation becomes higher at finer time resolution. This is a known fact for market variables (e.g. volume and volatility), while we document it for the first time at intraday scale also for browsing variables. The presence of a significant linear relation between the attention given to news articles (signed on the basis of the sentiment expressed in them) related to a given stock and the price return is mild. In particular the low fraction of companies rejecting the null hypothesis is compatible with the expected number of false positives due to multiple testing. Please refer to Materials Section for the detailed results of a multiple test based on the Bonferroni correction.

\subsection*{Dynamics of attention}
The time scale might in principle depend on the relevance of the news. As we have seen, not all news are equal in terms of the attention they receive from the users. To investigate this dependence, we study the dynamics of the number of clicks an article received after its publication. We compute the cumulative number of clicks received by a given news until a given minute after the publication. We perform this for all minutes in a week after the publication. We then normalize this cumulative time series by dividing it by the total amount of clicks received by the news. We construct ten groups of news based on the deciles of the total number of news they eventually receive, and we compute for each group the average cumulative sum of clicks.
\begin{figure}[th]
  \centering
  \includegraphics[width=0.75\textwidth]{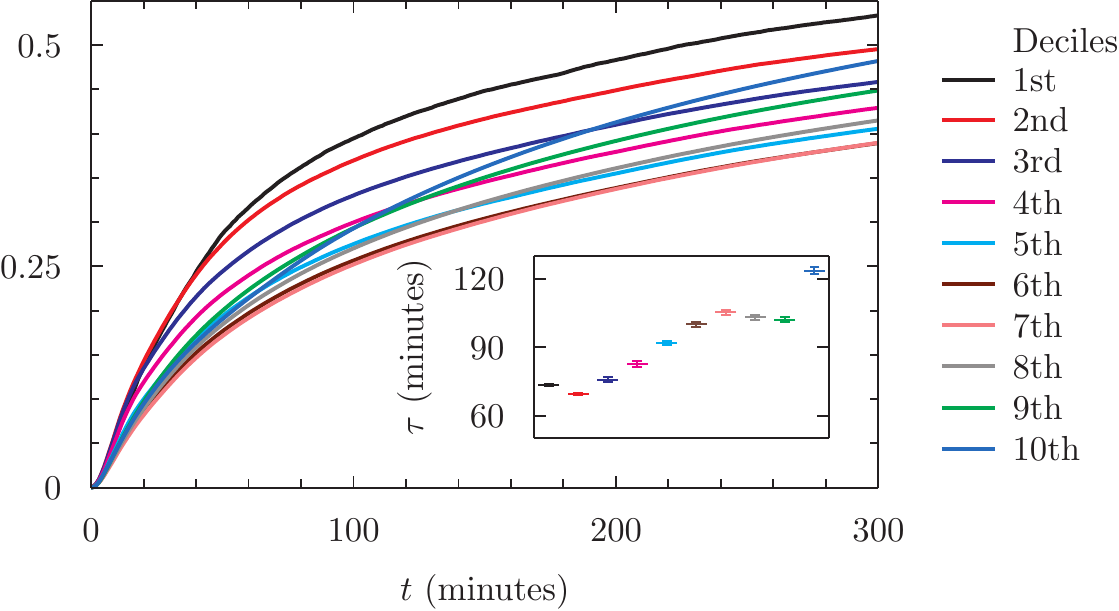}\\
  \caption{Time evolution of the cumulative number of clicks per news in a time interval of five hours after the publication. We normalize the cumulated amount by a constant which corresponds to the total number of clicks received by a single news during the first week after publication. The news are grouped in deciles according to the total number of clicks they have received until October 2013 and the curves represent average values. Inset: estimated values and standard errors of the attention time scale obtained by an exponential fit of the decile curves.} \label{dycdf}
\end{figure}
The result is shown in Figure~\ref{dycdf}. The inset reports for each decile the typical time scale of attention obtained by an exponential fit of the curves. Remarkably, the time scale of attention is an increasing function of the importance of the news (as measured by the total number of clicks). Irrelevant news are immediately recognized as such, while important news continue to receive attention well after their publication. In general, the time scale of the users' attention ranges between one and two hours after the publication, suggesting that this intraday time scale is probably the most appropriate to detect dependencies among financial variables and browsing activity.

\subsection*{Causality}
\begin{figure}[th]
  \centering
  \includegraphics[width=0.75\textwidth]{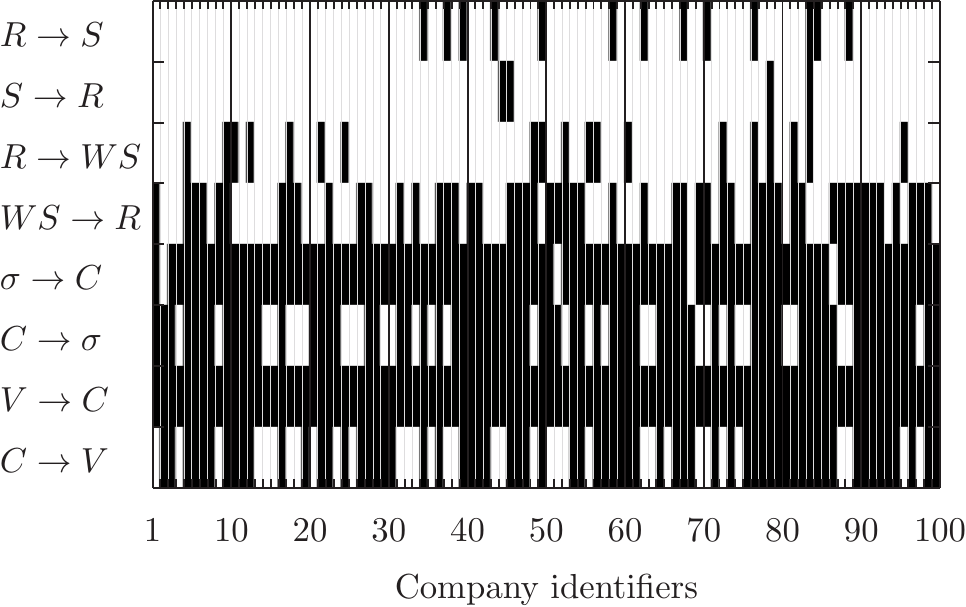}
  \caption{Granger Causality tests at hourly scale between de-seasonalized time series (x axis as in figure \ref{fcor1}). The white cells correspond to tests for which we do not reject the null hypothesis of no Granger causality at 5\% significance level. A black cell corresponds to a statistically significant Granger causality.}\label{gra1}
\end{figure}

The synchronous correlation is an important measure of dependence, but not necessarily a sign of causality. Thus we perform a causality analysis by applying Granger's test. We present the results of this analysis, for the 65-minute time horizon, in Figure~\ref{gra1}.
The x axis lists the companies as done in Figure~\ref{fcor1}, while the y axis labels the eight tests that we perform. Black cells correspond to rejection of the null hypothesis of no Granger causality, and the opposite for the white cells. When considering the non-negative variables ($V$, $C$, and $\sigma$) we observe strong causal relations. Specifically, in 65\% of the cases the clicking activity causes the trading volume and in 69\% of the cases it causes price volatility. The causality is very strong also in the opposite direction, i.e. volume and volatility cause click volume. This is probably  due in part to a reaction of users to anomalously high activity in the market (in terms of volume and/or volatility), while in part it might be a statistical effect due to the fact that all the three variables are very autocorrelated in time, creating strong Granger causal relations in both directions.

We obtain the most interesting and unexpected results when considering the signed variables ($R$, $S$, and $WS$). All these variables are weakly serially autocorrelated. When we consider the sentiment of the news ($S$, without the clicks), we find that only in 4\% of the cases $S$ causes returns, and in 13\% of the cases price return causes $S$. Especially the first value is expected under the null, since at 5\% confidence level we expect 5\% of false positive. This means that the simple sentiment of the news does not allow to forecast price returns at intraday (hourly) time scale. On the contrary, when we consider the clicks weighted by the sentiment of the news ($WS$), we find that in 53\% of the cases it allows predicting returns and only in 19\% of the cases the opposite occurs. In general, companies with more news have higher causality. Our conclusions are even more striking when correcting the test for multiple hypotheses, as done for the Spearman correlation case. The evidence of causality between price returns and unweighted sentiment of the news almost vanishes whereas the signal of causality between weighted sentiment and returns entirely survives. Interestingly, the evidence of causality in the opposite direction -- i.e. returns Granger-causing weighted sentiment -- weakens and an interesting asymmetric behavior between the two directions clearly emerges. In Materials Section we report the table with detailed results.

\subsection*{Weighting news by users' browsing behavior}
These results show that, on a hourly time scale, the simple news sentiment time series alone (i.e. the one without browsing activity) is not able to predict the price returns; instead, if we add the information provided by the browsing activity, we are then able to properly weigh the news (and its sentiment) by the importance the users give to it by clicking the page. Thus, we find the interesting result that the browsing activity combined with the sentiment analysis of the news increases significantly the level of prediction of price returns for the stocks.

\subsection*{Comparison with existing literature}
Most of the existing studies on sentiment and predictability of returns focus on daily or longer time scale. In order to compare properly our result with the existing literature, we present in Table \ref{gra3} the results of the above Granger tests  on a daily time scale. Table \ref{gra3} shows that, without the browsing activity, $S$ causes $R$ for 18\% of the companies and $R$ Granger-causes $S$ in 9\% of the cases. Thus there is now some predictability of sentiment, even if the number of companies is quite limited. This is consistent with the existing literature, which reports a weak daily predictability of returns by using  sentiment. It is important to note that by adding the browsing activity we can double the number of companies for which there is predictability. In fact, $WS$ Granger-causes $R$ for 37\% of the companies and 11\% in the opposite direction.
\begin{table}
  \caption{Number of companies for which we reject the null hypothesis of no Granger causality at 5\% confidence level.}
  \begin{tabular*}{\columnwidth}{@{\extracolsep{\fill}}lrr}
    Causality relation             & Hourly scale  &  Daily scale \cr
    \hline
    $S      \rightarrow R     $    &   4 &           18 \cr
    $R      \rightarrow S     $    &  13 &            9 \cr
    $R      \rightarrow WS    $    &  19 &           11 \cr
    $WS     \rightarrow R     $    &  53 &           37 \cr
    $V      \rightarrow C     $    & 100 &           97 \cr
    $C      \rightarrow V     $    &  65 &           52 \cr
    $C      \rightarrow \sigma$    &  69 &           52 \cr
    $\sigma \rightarrow C     $    &  96 &           16 \cr
    \hline
  \end{tabular*}
  \label{gra3}
\end{table}

\section{Materials}
\subsection{Financial time series} 
We consider only trading hours (i.e. from 9:30 AM to 4:00 PM) and trading days (i.e. business days at the New York Stock Exchange), and we disregard the trading events that occur out of this time window. From the financial data we create three time series: the first one is the traded volume for each minute for each company, the second one is the logarithmic returns, and the last time series is the absolute value of the logarithmic returns.
\begin{enumerate}
  \item \textbf{Trade volume (V) time series}. It consists of the traded volume of a given company at minute time scale. We build a time series at time scale $t$ by summing the traded volumes $v_\tau$ at the smallest time scale $\tau$ (in our case one minute). Then, the total volume $\overline{v_{t}} $ can be defined as follows:
    \begin{equation}
      \overline{v_{t}}= \sum_{\tau>t-1}^{t} v_{\tau}\,.
    \end{equation}
    Traded volumes at intra-day time $t$ are rescaled by a factor $\zeta^v_t$, which is computed as the average, over all days, of the volume at time $t$ normalized by the total daily volume. If $\overline{v_{d,t}}$ is the raw volume of day $d$ and intra-day time $t$, we define the rescaled time series as
    \begin{equation}
      V_{d,t}=\frac{\overline{v_{d,t}}}{\zeta^v_t}\,,
    \end{equation}	
    where
    \begin{equation}
      \zeta^v_{t}= \frac{1}{T} \sum_{ d^{'}} \frac{\overline{v_{d^{'},t}}}{\Lambda_{d^{'}}}\,,
    \end{equation}
    and
    \begin{equation}
      \Lambda _d=\sum_{t}\overline{v_{d,t}}\,,
    \end{equation}
    i.e. the total volume trades on day $d$, and $T$ is the total number of trading days.
  \item \textbf{Price return (R) time series}. We preliminary compute the logarithmic return, defined as
    \begin{equation}
      r_{t}= \log_{10}(\frac{p_{t}}{p_{t-1}})\,,
    \end{equation}
  where $p_{t}$ is the last recorded price in the interval $(t-1,t]$. Returns at intra-day time $t$ are rescaled by a factor $\zeta^r_t$, which is computed as the average, over all days, of absolute returns at time $t$ rescaled by the average volatility. More precisely, if $r_{d,t}$ is the raw return of day $d$ and intra-day time $t$, we define the rescaled time series as
  \begin{equation}
    R_{d,t}=\frac{r_{d,t}}{\zeta^r_t}\,,
  \end{equation}
  where
  \begin{equation}
    \zeta^r_{t}= \frac{1}{T} \sum_{ d^{'}} \frac{|r _{d^{'},t}|}{\Xi_{d^{'}}}\,,
  \end{equation}
  and
  \begin{equation}
    \Xi_d= \mathrm{mean}(|r_{d,t}|)\,,
  \end{equation}
  where we compute the mean value for all $t$ in a day $d$.
\item \textbf{Volatility ($\sigma$) time series}. We employ as a simple proxy for the de-seasonalized intra-day volatility the absolute value of the rescaled return
  \begin{equation}
    \sigma_{t}= |R_{t}|\,.
  \end{equation}
  Notice that the volatility $\sigma_{t}$ is already de-seasonalized, because of the definition of $R$
\end{enumerate}

\subsection{Sentiment analysis}
We discuss here how we tagged the news and how we performed the sentiment analysis. The news data contained in the log of Yahoo! Finance were marked with relevant stocks, identified through two different tagging methods. A first type of annotations was provided by a team of editors, who manually assessed the articles published on the Yahoo! Finance website and marked them with companies which the content of the article was relevant for. In addition to that, relevant companies were also identified by applying a proprietary entity-recognition tool. We discarded all the articles that were tagged with more than 4 companies. This threshold was chosen heuristically in a preliminary assessment, where we verified that articles tagged with 5 or more companies generally corresponds to aggregated periodic reports, which mention long lists of stocks without being really specific about any of them.

Among the obtained articles, we further processed those tagged with more than one company, introducing some additional filters to make sure we would consider the news for a company when the content of the article was really relevant to it, and not in the case where a company was only tagged due to a casual mention in the text. Such additional filters consisted in only retaining the stocks that were mentioned either in the title or in the first paragraph of the article, which typically contains a summary of the key concepts discussed in the article. Each article then contributed to the time series of all the stocks it was tagged for.

As explained in the main text, we have used {\em SentiStrength} adding a list of sentiment keywords of special interest and significance for the financial domain. We have tested the classification in good, bad, or neutral news using the general dictionary. In 84\% of the cases the classification is the same by using both dictionary. This result indicates that our classification and the consequent analysis is pretty robust to the choice of dictionary. Interestingly, 17\% of the news classified as neutral by the general dictionary are classified as positive or negative by using the financial dictionary, while only 8\% of the news classified as neutral by the financial dictionary are classified as positive or negative when using the general dictionary. This suggests that the use of a financial dictionary sharpens the capability of giving a positive or negative sign to a news.

\subsection{Click and sentiment time series}
Consistently with the financial time series, we consider only trading time and trading days. This implies that we neglect the clicks that occur out of this time window, because we are only interested in the click behaviour during the trading hours evolution of markets. From the click-through history of each news we create two time series: the first one is the total amount of clicks for each minute for each company, and the second one is the number of clicks multiplied by the sentiment score of the associated news.

\begin{enumerate}
  \item \textbf{Click (C) time series}. Starting from the number of clicks of each news at the smallest time scale $\tau$ (in our case one minute), we build
    a time series at the time scale $t$ by aggregating the number of clicks of all the news of a given company. So if we denote by $N$ the number of news of a company, and by $c_{\tau}^i$ the number of clicks for news $i$ at scale $\tau$, the total volume $\overline{c}_t$ can be defined as
    \begin{equation}
      \overline{c_{t}}= \sum_{\tau>t-1}^{t} \sum_{i=1}^{N} c_{\tau}^{i}\,.
    \end{equation}
    The news that are not viewed in the time interval have zero clicks. We filter out the intra-day pattern from clicks by means of a simple methodology. Clicks at intra-day time $t$ are rescaled by a factor $\zeta^c_t$, which is computed as the average, over all days, of the click volume at time $t$ normalized by the total number of daily clicks. More precisely, if $\overline{c_{d,t}}$ is the raw click volume of day $d$ and intra-day time $t$, we define the time series of rescaled clicks as:
    \begin{equation}
      C_{d,t}=\frac{\overline{c}_{d,t}}{\zeta^c_t}\,,
    \end{equation}
    where
    \begin{equation}
      \zeta^c_{t}= \frac{1}{T} \sum_{ d^{'}} \frac{\overline{c}_{d^{'},t}}{\Gamma_{d^{'}}}\,,
    \end{equation}
    and
    \begin{equation}
      \Gamma_d=\sum_{t} \overline{c}_{d,t}\,,
    \end{equation}
    with $\Gamma_d$ the total number of clicks in a day.
  \item \textbf{Sentiment (S) time series}. To construct this time series we consider the sentiment of the headlines of the news. With the same notation previously used
    \begin{equation}
      S_{t}= \sum_{\tau>t-1}^{t} \sum_{i=1}^{N} s_{\tau}^{i}\,,
    \end{equation}
    where $s_{\tau}^{i}$ is the sign $(-1,0,1)$ of the sentiment of a news headline published at time $\tau$.
  \item \textbf{Weighted Sentiment (WS) time series}. We multiply the click count of each news by its sentiment score. With the same notation of the click time series, we have:
    \begin{equation}
      \overline{WS}_{t}= \sum_{\tau>t-1}^{t} \sum_{i=1}^{N} c_{\tau}^{i} s_{\tau}^{i}\,.
    \end{equation}
\end{enumerate}
This way, we weight the sign of every news by the level of attention that the news has received. We remark that for each trading day we consider all the clicks of all the news clicked in that day, not only the news that have been released in that day. Then, we define the weighted sentiment as
\begin{equation}
  WS_{t}= \mathrm{sign}(\overline{WS}_{t})C_{t}\,.
\end{equation}

\subsection{Estimation of power law exponents}
We estimate the power law tail exponent of the distribution of the number of clicks per news by employing the R package PoweRlaw developed and maintained by Colin Gillespie, and described in~\cite{Clauset_etal:2009}. See Table~\ref{100stockspl} for numerical details.
\begin{table}
  \caption{Tail exponent $\alpha$ and lower bound $x_\mathrm{min}>0$ with standard errors $\epsilon_\alpha$ and $\epsilon_{x_\mathrm{min}}$ estimated from 100 highly capitalized stocks traded in the US equity markets. Following the procedure detailed in~\cite{Clauset_etal:2009}, we estimate the probability distribution associated with integer numbers larger than $x_\mathrm{min}$ whose expression reads to $p(x) = x^{-1-\alpha}/\zeta(1+\alpha,x_\mathrm{min})$. The normalizing constant corresponds to the Hurwitz zeta function $\zeta$. We report in bold the top 10 assets, as measured by the absolute number of associated news.}
  {\scriptsize
    \begin{tabular*}{\columnwidth}{@{\extracolsep{\fill}}lrrrrlrrrr}
      & $\alpha$  &  $\epsilon_\alpha$ & $x_\mathrm{min}$ & $\epsilon_{x_\mathrm{min}}$ &  & $\alpha$  &  $\epsilon_\alpha$ & $x_\mathrm{min}$ & $\epsilon_{x_\mathrm{min}}$ \cr
      \hline
      AA   & 1.05 & 0.09 &  482 &  114 & \textbf{JPM}  & \textbf{0.99} & \textbf{0.04} &  \textbf{665} &  \textbf{205} \cr
      \textbf{AAPL} & \textbf{0.97} & \textbf{0.05} & \textbf{2881} & 1\textbf{451} & KBH  & 1.33 & 0.20 &  402 &   98 \cr
      ABT  & 1.39 & 0.33 &  530 &  261 & KO   & 0.96 & 0.09 &  565 &  242 \cr
      ACN  & 1.84 & 0.37 & 1074 &  313 & LCC  & 0.95 & 0.08 &  722 &  371 \cr
      AET  & 1.21 & 0.13 &  261 &   54 & LEN  & 0.86 & 0.19 &  256 &  197 \cr
      AIG  & 1.16 & 0.13 &  854 &  193 & LMT  & 1.41 & 0.14 &  808 &  159 \cr
      AMGN & 1.60 & 0.33 & 1596 &  445 & LNKD & 0.90 & 0.13 &  735 &  297 \cr
      AMZN & 1.06 & 0.05 &  969 &  206 & LUV  & 0.82 & 0.10 &  617 &  282 \cr
      ATVI & 1.47 & 0.23 &  863 &  192 & MA   & 0.97 & 0.13 &  335 &  213 \cr
      AXP  & 0.89 & 0.07 &  447 &   93 & MAR  & 1.34 & 0.17 &  318 &   90 \cr
      \textbf{BAC}  & \textbf{0.95} & \textbf{0.05} &  \textbf{807} &  \textbf{200} & MCD  & 0.71 & 0.06 &  462 &  305 \cr
      \textbf{BA}   & \textbf{1.01} & \textbf{0.05} &  \textbf{802} &  \textbf{250} & M    & 0.86 & 0.09 &  384 &  200 \cr
      BBY  & 0.74 & 0.13 &  957 &  526 & MRK  & 1.53 & 0.20 &  966 &  213 \cr
      BIIB & 1.88 & 0.41 & 1081 &  251 & MSCI & 1.12 & 0.09 &  205 &   62 \cr
      BLK  & 1.20 & 0.12 &  547 &  122 & \textbf{MS  } & \textbf{1.15} & \textbf{0.09} &  \textbf{615} &  \textbf{457} \cr
      CAT  & 1.14 & 0.11 &  834 &  224 & \textbf{MSFT} & \textbf{1.06} & \textbf{0.05} & \textbf{2247} &  \textbf{507} \cr
      CBS  & 0.82 & 0.07 &  169 &   75 & NDAQ & 1.24 & 0.08 &  213 &   42 \cr
      \textbf{C}    & \textbf{1.10} & \textbf{0.06} &  \textbf{627} &  \textbf{256} & NFLX & 0.94 & 0.07 &  613 &  141 \cr
      CHK  & 1.72 & 0.18 & 1479 &  215 & NKE  & 1.09 & 0.14 &  566 &  143 \cr
      CMCSA& 1.41 & 0.24 & 1186 &  357 & NOC  & 1.46 & 0.22 &  894 &  228 \cr
      COP  & 1.76 & 0.24 & 1708 &  303 & NWSA & 0.89 & 0.06 &  234 &  117 \cr
      COST & 0.79 & 0.10 &  448 &  372 & NYX  & 0.98 & 0.18 &  176 &  192 \cr
      CSCO & 1.32 & 0.11 & 1290 &  271 & ORCL & 0.95 & 0.20 &  523 &  422 \cr
      CVX  & 1.37 & 0.15 &  782 &  180 & PEP  & 0.92 & 0.13 &  565 &  220 \cr
      DAL  & 1.01 & 0.12 &  959 &  220 & PFE  & 1.54 & 0.14 & 1306 &  250 \cr
      DELL & 1.00 & 0.07 &  944 &  246 & PG   & 1.02 & 0.11 &  809 &  327 \cr
      DIS  & 0.72 & 0.04 &  283 &  198 & QCOM & 1.44 & 0.23 & 1748 &  363 \cr
      DISH & 1.05 & 0.17 &  515 &  391 & RTN  & 1.83 & 0.47 & 1379 &  425 \cr
      DOW  & 1.21 & 0.13 &  818 &  313 & SBUX & 0.80 & 0.07 &  453 &  236 \cr
      DTV  & 1.11 & 0.13 &  438 &  132 & S    & 1.09 & 0.14 &  684 &  424 \cr
      EBAY & 0.91 & 0.11 & 1612 &  459 & TRI  & 0.89 & 0.04 &  122 &   62 \cr
      \textbf{FB}   & \textbf{0.95} & \textbf{0.08} & \textbf{1211} & \textbf{1077} & TSLA & 0.98 & 0.11 & 3024 &  850 \cr
      F    & 0.96 & 0.05 & 1188 &  254 & TWC  & 1.06 & 0.09 &  403 &  176 \cr
      FRT  & 1.19 & 0.08 &  412 &   87 & TWX  & 0.84 & 0.08 &  260 &  119 \cr
      GE   & 0.98 & 0.08 &  587 &  236 & TXN  & 1.55 & 0.42 & 1060 &  475 \cr
      GILD & 1.87 & 0.43 & 1300 &  294 & UAL  & 1.00 & 0.12 & 1153 &  659 \cr
      GM   & 0.84 & 0.05 &  863 &  296 & UNH  & 1.21 & 0.15 &  378 &  147 \cr
      \textbf{GOOG} & \textbf{0.92} & \textbf{0.03} & \textbf{1786} &  \textbf{450} & UPS  & 1.93 & 0.49 & 2499 &  721 \cr
      GPS  & 1.28 & 0.17 &  390 &  158 & UTX  & 1.49 & 0.20 &  733 &  209 \cr
      GRPN & 0.93 & 0.13 &  644 &  193 & V    & 1.17 & 0.13 &  505 &  243 \cr
      \textbf{GS}   & \textbf{0.89} & \textbf{0.05} &  \textbf{462} &  \textbf{143} & VMW  & 1.50 & 0.45 & 1003 &  547 \cr
      HAL  & 1.56 & 0.18 &  880 &  213 & VOD  & 1.87 & 0.42 & 1850 &  711 \cr
      HD   & 0.79 & 0.07 &  456 &  169 & VZ   & 1.11 & 0.10 & 1001 &  359 \cr
      HON  & 1.18 & 0.17 &  632 &  296 & WAG  & 1.06 & 0.23 &  512 &  322 \cr
      HPQ  & 1.13 & 0.11 &  786 &  274 & WFC  & 0.95 & 0.05 &  874 &  271 \cr
      IBM  & 1.11 & 0.11 & 1014 &  283 & WMT  & 0.67 & 0.06 & 2073 &  517 \cr
      INTC & 1.35 & 0.32 & 1777 &  738 & XOM  & 1.29 & 0.12 &  802 &  147 \cr
      IT   & 1.32 & 0.13 &  410 &   98 & YHOO & 1.15 & 0.12 & 2546 &  605 \cr
      JCP  & 0.75 & 0.07 & 2827 &  959 & YUM  & 0.90 & 0.10 &  483 &  225 \cr
      JNJ  & 1.51 & 0.19 & 1058 &  174 & ZNGA & 1.38 & 0.16 & 1578 &  338 \cr
      \hline
    \end{tabular*}
    \label{100stockspl}
  }
\end{table}

\subsection{Multiple hypotheses testing}
In Tables~\ref{tab:Spearman-Bonferroni} and~\ref{tab:Granger-Bonferroni} we report the results for tests of zero Spearman correlation and zero Granger causality, respectively, under the very conservative correction proposed by Bonferroni. If $N_t$ tests are performed and the desired significance is $p$ (in our case $5\%$, then only the tests with a p-value smaller than $p/N_t$ are rejected. Since we perform $100$ tests, our corrected p-value is $0.05\%$.
\begin{table}
  \caption{Percentage of companies for which we reject the null hypothesis of zero Spearman correlation at 0.05\% confidence level (Bonferroni's correction).}
  \label{tab:Spearman-Bonferroni}
  \begin{tabular*}{\columnwidth}{@{\extracolsep{\fill}}lrrr}
    Time interval (minutes) &  $\rho(WS,R)$ &  $\rho(C,\sigma)$ &  $\rho(C,V)$ \cr
    \hline
    1        &      0 &       73 &      92 \cr
    10       &      0 &       48 &      81 \cr
    30       &      0 &       24 &      76 \cr
    65       &      0 &        9 &      66 \cr
    130      &      0 &        6 &      55 \cr
    \hline
  \end{tabular*}
\end{table}
\begin{table}
  \caption{Number of companies for which we reject the null hypothesis of no Granger causality at 0.05\% confidence level (Bonferroni's correction).}
  \begin{tabular*}{\columnwidth}{@{\extracolsep{\fill}}lr}
    Causality relation             & Hourly scale \cr
    \hline
    $S      \rightarrow R     $    &   1 \cr
    $R      \rightarrow S     $    &   0 \cr
    $R      \rightarrow WS    $    &   5 \cr
    $WS     \rightarrow R     $    &  49 \cr
    $V      \rightarrow C     $    &  94 \cr
    $C      \rightarrow V     $    &  52 \cr
    $C      \rightarrow \sigma$    &  60 \cr
    $\sigma \rightarrow C     $    &  77 \cr
    \hline
  \end{tabular*}
  \label{tab:Granger-Bonferroni}
\end{table}

\section{Discussion} \label{sec:IV}
The semantic analysis of the news on a specific company is known to have a small predictive power on the future price movements. At the light of our findings, we argue that this effect could be related to the distribution in the attention that the news receive, as clearly emerge in Figure 1: its scale-free behaviour reflects the extreme heterogeneity in the information they convey and the surprise they generate in the readers.

Our in-sample analysis shows that by adding the clicking activity of the web users, we can greatly increase the predictive power of the news for the price returns. This occurs because the time series built with only the sentiment of the news gives the same weight to all the news. In this way even irrelevant news are considered, adding noise to the sentiment time series and reducing the predictive power of the signal. Adding the browsing activity means giving a meaningful weight to each news according to its importance, as measured by the attention it receives by the users, and it enhances the forecast ability of our approach.

The approach to collective evaluation that we proposed in this paper can be useful in many other non financial contexts, since the overflow of information is a common aspect in our lives. In the financial domain, a natural extension of the present work concerns market instabilities and crashes. The analysis presented here is in fact unconditional, i.e. it does not target large price movements or, more generally, abnormal returns. From our societal perspective it would be extremely valuable to have a collective evaluation system, like the one presented here, capable of sifting the relevant information from the pool of data, news, blogs, etc, and to provide early warning indicators of large price movements. Since we have reported evidences in favour of return predictability at intraday time scale -- especially at hourly scale -- this approach could be also used for real-time indicators, as well as for high-frequency instabilities and systemic price cojumps \cite{bormetti2013modelling,calcagnile2015}, which are becoming increasingly more frequent in our highly automated financial world.

\section{Acknowledgments}
The authors warmly thank Lucio Calcagnile for the valuable technical support during the final stage of this work. GR and GC  acknowledge support from FET Open Project SIMPOL ``Financial Systems Simulation and Policy Modelling'' nr. 610704, IP Project MULTIPLEX nr 317532 and Progetto di Interesse CNR ``CrisisLab''. GC acknowledges also partial support  from the U.S. Department of the Defense, Defense Threat Reduction Agency, grant HDTRA1-11-1-0048. FL acknowledges support from the FP7/2007-2013 under grant agreement CRISIS-ICT-2011-288501. Data have been made available by Yahoo! inc and Quantlab groups.


\begin{thebibliography}{43}
    \expandafter\ifx\csname url\endcsname\relax
    \def\url#1{\texttt{#1}}\fi
    \expandafter\ifx\csname urlprefix\endcsname\relax\def\urlprefix{URL }\fi
    \providecommand{\bibinfo}[2]{#2}
    \providecommand{\eprint}[2][]{\url{#2}}

  \bibitem{king2011ensuring}
    \bibinfo{author}{King, G.}
    \newblock \bibinfo{title}{{Ensuring the data-rich future of the social
    sciences}}.
    \newblock \emph{\bibinfo{journal}{Science}} \textbf{\bibinfo{volume}{331}},
    \bibinfo{pages}{719--721} (\bibinfo{year}{2011}).

  \bibitem{vespignani2009predicting}
    \bibinfo{author}{Vespignani, A.}
    \newblock \bibinfo{title}{{Predicting the behavior of techno-social systems}}.
    \newblock \emph{\bibinfo{journal}{Science}} \textbf{\bibinfo{volume}{325}},
    \bibinfo{pages}{425--428} (\bibinfo{year}{2009}).

  \bibitem{bonanno2004networks}
    \bibinfo{author}{Bonanno, G.} \emph{et~al.}
    \newblock \bibinfo{title}{Networks of equities in financial markets}.
    \newblock \emph{\bibinfo{journal}{The European Physical Journal B-Condensed
    Matter and Complex Systems}} \textbf{\bibinfo{volume}{38}},
    \bibinfo{pages}{363--371} (\bibinfo{year}{2004}).

  \bibitem{caldarelli2007scale}
    \bibinfo{author}{Caldarelli, G.}
    \newblock \emph{\bibinfo{title}{{Scale-Free Networks: complex webs in nature
    and technology}}} (\bibinfo{publisher}{Oxford University Press},
    \bibinfo{address}{Oxford}, \bibinfo{year}{2007}).

  \bibitem{tumminello2005tool}
    \bibinfo{author}{Tumminello, M.}, \bibinfo{author}{Aste, T.},
    \bibinfo{author}{{Di Matteo}, T.} \& \bibinfo{author}{Mantegna, R.~N.}
    \newblock \bibinfo{title}{{A tool for filtering information in complex
    systems.}}
    \newblock \emph{\bibinfo{journal}{Proceedings of the National Academy of
    Sciences of the United States of America}} \textbf{\bibinfo{volume}{102}},
    \bibinfo{pages}{10421--10426} (\bibinfo{year}{2005}).

  \bibitem{achrekar2011predicting}
    \bibinfo{author}{Achrekar, H.}, \bibinfo{author}{Gandhe, A.},
    \bibinfo{author}{Lazarus, R.}, \bibinfo{author}{Yu, S.-H.} \&
    \bibinfo{author}{Liu, B.}
    \newblock \bibinfo{title}{{Predicting flu trends using twitter data}}.
    \newblock In \emph{\bibinfo{booktitle}{Computer Communications Workshops
    (INFOCOM WKSHPS), 2011 IEEE Conference on}}, \bibinfo{pages}{702--707}
    (\bibinfo{organization}{IEEE}, \bibinfo{year}{2011}).

  \bibitem{tizzoni2012real}
    \bibinfo{author}{Tizzoni, M.} \emph{et~al.}
    \newblock \bibinfo{title}{{Real-time numerical forecast of global epidemic
    spreading: case study of 2009 A/H1N1pdm.}}
    \newblock \emph{\bibinfo{journal}{BMC medicine}} \textbf{\bibinfo{volume}{10}},
    \bibinfo{pages}{165} (\bibinfo{year}{2012}).

  \bibitem{caldarelli2014multilevel}
    \bibinfo{author}{Caldarelli, G.} \emph{et~al.}
    \newblock \bibinfo{title}{{A multi-level geographical study of Italian
    political elections from Twitter data.}}
    \newblock \emph{\bibinfo{journal}{PloS one}} \textbf{\bibinfo{volume}{9}},
    \bibinfo{pages}{e95809} (\bibinfo{year}{2014}).

  \bibitem{malkiel1970efficient}
    \bibinfo{author}{Malkiel, B.~G.} \& \bibinfo{author}{Fama, E.~F.}
    \newblock \bibinfo{title}{{Efficient capital markets: A review of theory and
    empirical work}}.
    \newblock \emph{\bibinfo{journal}{The Journal of Finance}}
    \textbf{\bibinfo{volume}{25}}, \bibinfo{pages}{383--417}
    (\bibinfo{year}{1970}).

  \bibitem{thelwall2010sentiment}
    \bibinfo{author}{Thelwall, M.}, \bibinfo{author}{Buckley, K.},
    \bibinfo{author}{Paltoglou, G.}, \bibinfo{author}{Cai, D.} \&
    \bibinfo{author}{Kappas, A.}
    \newblock \bibinfo{title}{{Sentiment strength detection in short informal
    text}}.
    \newblock \emph{\bibinfo{journal}{Journal of the American Society for
    Information Science and Technology}} \textbf{\bibinfo{volume}{61}},
    \bibinfo{pages}{2544--2558} (\bibinfo{year}{2010}).

  \bibitem{loughran2011liability}
    \bibinfo{author}{Loughran, T.} \& \bibinfo{author}{McDonald, B.}
    \newblock \bibinfo{title}{{When is a liability not a liability? Textual
    analysis, dictionaries, and 10-Ks}}.
    \newblock \emph{\bibinfo{journal}{The Journal of Finance}}
    \textbf{\bibinfo{volume}{66}}, \bibinfo{pages}{35--65}
    (\bibinfo{year}{2011}).

  \bibitem{chan2003stock}
    \bibinfo{author}{Chan, W.~S.}
    \newblock \bibinfo{title}{{Stock price reaction to news and no-news: drift and
    reversal after headlines}}.
    \newblock \emph{\bibinfo{journal}{Journal of Financial Economics}}
    \textbf{\bibinfo{volume}{70}}, \bibinfo{pages}{223--260}
    (\bibinfo{year}{2003}).

  \bibitem{tetlock2007giving}
    \bibinfo{author}{Tetlock, P.~C.}
    \newblock \bibinfo{title}{{Giving content to investor sentiment: The role of
    media in the stock market}}.
    \newblock \emph{\bibinfo{journal}{The Journal of Finance}}
    \textbf{\bibinfo{volume}{62}}, \bibinfo{pages}{1139--1168}
    (\bibinfo{year}{2007}).

  \bibitem{mao2011predicting}
    \bibinfo{author}{Mao, H.}, \bibinfo{author}{Counts, S.} \&
    \bibinfo{author}{Bollen, J.}
    \newblock \bibinfo{title}{{Predicting financial markets: Comparing survey,
    news, twitter and search engine data}}.
    \newblock \emph{\bibinfo{journal}{arXiv preprint arXiv:1112.1051}}
    (\bibinfo{year}{2011}).

  \bibitem{lillo2012how}
    \bibinfo{author}{Lillo, F.}, \bibinfo{author}{Miccich\`{e}, S.},
    \bibinfo{author}{Tumminello, M.}, \bibinfo{author}{Piilo, J.} \&
    \bibinfo{author}{Mantegna, R.~N.}
    \newblock \bibinfo{title}{{How news affect the trading behavior of different
    categories of investors in a financial market}}.
    \newblock \emph{\bibinfo{journal}{Quantitative Finance}}
    (\bibinfo{year}{2014}).
    \newblock \bibinfo{note}{DOI: 10.1080/14697688.2014.931593}.

  \bibitem{bordino2014stock}
    \bibinfo{author}{Bordino, I.}, \bibinfo{author}{Kourtellis, N.},
    \bibinfo{author}{Laptev, N.} \& \bibinfo{author}{Billawala, Y.}
    \newblock \bibinfo{title}{Stock trade volume prediction with yahoo finance user
    browsing behavior}.
    \newblock In \emph{\bibinfo{booktitle}{Data Engineering (ICDE), 2014 IEEE 30th
    International Conference on}}, \bibinfo{pages}{1168--1173}
    (\bibinfo{year}{2014}).

  \bibitem{wang2013financial}
    \bibinfo{author}{Wang, C.},  \bibinfo{author}{Tsai, M.},  \bibinfo{author}{Liu, T.},   \& \bibinfo{author}{Chang, C.}
    \newblock \bibinfo{title}{Financial Sentiment Analysis for Risk Prediction.}
    \newblock In \emph{\bibinfo{booktitle}{Proceedings of the Sixth International Joint Conference on Natural Language Processing}},
    \bibinfo{pages}{802--808},
    (\bibinfo{year}{2013}).

  \bibitem{reis2015breaking}
    \bibinfo{author}{Reis, J. C.},  \bibinfo{author}{Benvenuto, F.},  \bibinfo{author}{Vaz de Melo, P.},   \bibinfo{author}{Prates, O.},  \bibinfo{author}{Kwak, H.},   \& \bibinfo{author}{An, J.}
    \newblock \bibinfo{title}{Breaking the News: First Impressions Matter on Online News}
    \newblock In \emph{\bibinfo{booktitle}{ICWSM'15: Proceedings of The International Conference on Weblogs and Social Media, 2015}},
    (\bibinfo{year}{2015}).

  \bibitem{martinez2012semantic}
    \bibinfo{author}{Ruiz-Martinez, J.M.},  \bibinfo{author}{Valencia-Garcia, R.},  \& \bibinfo{author}{Garcia-Sanchez, F.}
    \newblock \bibinfo{title}{Semantic-Based Sentiment analysis in financial news.}
    \newblock In \emph{\bibinfo{booktitle}{Proceedings of the  First International Workshop on Finance and Economics on the Semantic Web (FEOSW 2012) in conjunction with 9th Extended Semantic Web Conference (ESWC 2012),}}
    (\bibinfo{year}{2012}).

  \bibitem{bank2011google}
    \bibinfo{author}{Bank, M.}, \bibinfo{author}{Larch, M.} \&
    \bibinfo{author}{Peter, G.}
    \newblock \bibinfo{title}{Google search volume and its influence on liquidity
    and returns of german stocks}.
    \newblock \emph{\bibinfo{journal}{Financial markets and portfolio management}}
    \textbf{\bibinfo{volume}{25}}, \bibinfo{pages}{239--264}
    (\bibinfo{year}{2011}).

  \bibitem{bordino2012web}
    \bibinfo{author}{Bordino, I.} \emph{et~al.}
    \newblock \bibinfo{title}{{Web search queries can predict stock market
    volumes}}.
    \newblock \emph{\bibinfo{journal}{PloS One}} \textbf{\bibinfo{volume}{7}},
    \bibinfo{pages}{e40014} (\bibinfo{year}{2012}).

  \bibitem{preis2010complex}
    \bibinfo{author}{Preis, T.}, \bibinfo{author}{Reith, D.} \&
    \bibinfo{author}{Stanley, H.~E.}
    \newblock \bibinfo{title}{{Complex dynamics of our economic life on different
    scales: insights from search engine query data}}.
    \newblock \emph{\bibinfo{journal}{Philosophical Transactions of the Royal
    Society A: Mathematical, Physical and Engineering Sciences}}
    \textbf{\bibinfo{volume}{368}}, \bibinfo{pages}{5707--5719}
    (\bibinfo{year}{2010}).

  \bibitem{kristoufek2013can}
    \bibinfo{author}{Kristoufek, L.}
    \newblock \bibinfo{title}{Can google trends search queries contribute to risk
    diversification?}
    \newblock \emph{\bibinfo{journal}{Scientific Reports}}
    \textbf{\bibinfo{volume}{3}} (\bibinfo{year}{2013}).

  \bibitem{vlastakis2012information}
    \bibinfo{author}{Vlastakis, N.} \& \bibinfo{author}{Markellos, R.~N.}
    \newblock \bibinfo{title}{Information demand and stock market volatility}.
    \newblock \emph{\bibinfo{journal}{Journal of Banking \& Finance}}
    \textbf{\bibinfo{volume}{36}}, \bibinfo{pages}{1808--1821}
    (\bibinfo{year}{2012}).

  \bibitem{zhangetal2013em}
    \bibinfo{author}{Zhang, W.}, \bibinfo{author}{Shen, D.}, \bibinfo{author}{Zhang, Y.} \& \bibinfo{author}{Xiong, X.}
    \newblock \bibinfo{title}{Open source information, investor attention, and asset pricing}.
    \newblock \emph{\bibinfo{journal}{Economic Modelling}}
    \textbf{\bibinfo{volume}{33}}, \bibinfo{pages}{613--619}
    (\bibinfo{year}{2013}).

  \bibitem{curme2014quantifying}
    \bibinfo{author}{Curme, C.}, \bibinfo{author}{Preis, T.},
    \bibinfo{author}{Stanley, H.~E.} \& \bibinfo{author}{Moat, H.~S.}
    \newblock \bibinfo{title}{{Quantifying the semantics of search behavior before
    stock market moves}}.
    \newblock \emph{\bibinfo{journal}{Proceedings of the National Academy of
    Sciences}} \textbf{\bibinfo{volume}{111}}, \bibinfo{pages}{11600--11605}
    (\bibinfo{year}{2014}).

  \bibitem{cutler1998moves}
    \bibinfo{author}{Cutler, D.~M.}, \bibinfo{author}{Poterba, J.~M.} \&
    \bibinfo{author}{Summers, L.~H.}
    \newblock \bibinfo{title}{{What moves stock prices?}}
    \newblock \emph{\bibinfo{journal}{Journal of Portfolio Management}}
    \textbf{\bibinfo{volume}{15}}, \bibinfo{pages}{4--12} (\bibinfo{year}{1989}).

  \bibitem{vega2006stock}
    \bibinfo{author}{Vega, C.}
    \newblock \bibinfo{title}{{Stock price reaction to public and private
    information}}.
    \newblock \emph{\bibinfo{journal}{Journal of Financial Economics}}
    \textbf{\bibinfo{volume}{82}}, \bibinfo{pages}{103--133}
    (\bibinfo{year}{2006}).

  \bibitem{tetlock2008more}
    \bibinfo{author}{Tetlock, P.~C.}, \bibinfo{author}{Saar-Tsechansky, M.} \&
    \bibinfo{author}{Macskassy, S.}
    \newblock \bibinfo{title}{{More than words: Quantifying language to measure
    firms' fundamentals}}.
    \newblock \emph{\bibinfo{journal}{The Journal of Finance}}
    \textbf{\bibinfo{volume}{63}}, \bibinfo{pages}{1437--1467}
    (\bibinfo{year}{2008}).

  \bibitem{schumaker2009textual}
    \bibinfo{author}{Schumaker, R.~P.} \& \bibinfo{author}{Chen, H.}
    \newblock \bibinfo{title}{Textual analysis of stock market prediction using
      breaking financial news: The {AZF}in text system}.
      \newblock \emph{\bibinfo{journal}{ACM Transactions on Information Systems
      (TOIS)}} \textbf{\bibinfo{volume}{27}}, \bibinfo{pages}{12}
      (\bibinfo{year}{2009}).

    \bibitem{engelberg2012shorts}
      \bibinfo{author}{Engelberg, J.~E.}, \bibinfo{author}{Reed, A.~V.} \&
      \bibinfo{author}{Ringgenberg, M.~C.}
      \newblock \bibinfo{title}{{How are shorts informed? Short sellers, news, and
      information processing}}.
      \newblock \emph{\bibinfo{journal}{Journal of Financial Economics}}
      \textbf{\bibinfo{volume}{105}}, \bibinfo{pages}{260--278}
      (\bibinfo{year}{2012}).

    \bibitem{alanyali2013quantifying}
      \bibinfo{author}{Alanyali, M.}, \bibinfo{author}{Moat, H.~S.} \&
      \bibinfo{author}{Preis, T.}
      \newblock \bibinfo{title}{Quantifying the relationship between financial news
      and the stock market}.
      \newblock \emph{\bibinfo{journal}{Scientific Reports}}
      \textbf{\bibinfo{volume}{3}} (\bibinfo{year}{2013}).

    \bibitem{zhang2014internet}
      \bibinfo{author}{Zhang, Y.} \emph{et~al.}
      \newblock \bibinfo{title}{Internet information arrival and volatility of sme
      price index}.
      \newblock \emph{\bibinfo{journal}{Physica A: Statistical Mechanics and its
      Applications}} \textbf{\bibinfo{volume}{399}}, \bibinfo{pages}{70--74}
      (\bibinfo{year}{2014}).

    \bibitem{birz2011effect}
      \bibinfo{author}{Birz, G.} \& \bibinfo{author}{{Lott Jr}, J.~R.}
      \newblock \bibinfo{title}{{The effect of macroeconomic news on stock returns:
      New evidence from newspaper coverage}}.
      \newblock \emph{\bibinfo{journal}{Journal of Banking \& Finance}}
      \textbf{\bibinfo{volume}{35}}, \bibinfo{pages}{2791--2800}
      (\bibinfo{year}{2011}).

    \bibitem{gross2011machines}
      \bibinfo{author}{Gross-Klussmann, A.} \& \bibinfo{author}{Hautsch, N.}
      \newblock \bibinfo{title}{{When machines read the news: Using automated text
      analytics to quantify high frequency news-implied market reactions}}.
      \newblock \emph{\bibinfo{journal}{Journal of Empirical Finance}}
      \textbf{\bibinfo{volume}{18}}, \bibinfo{pages}{321--340}
      (\bibinfo{year}{2011}).

    \bibitem{da2011search}
      \bibinfo{author}{Da, Z.}, \bibinfo{author}{Engelberg, J.} \&
      \bibinfo{author}{Gao, P.}
      \newblock \bibinfo{title}{{In search of attention}}.
      \newblock \emph{\bibinfo{journal}{The Journal of Finance}}
      \textbf{\bibinfo{volume}{66}}, \bibinfo{pages}{1461--1499}
      (\bibinfo{year}{2011}).

    \bibitem{de2008can}
      \bibinfo{author}{De~Choudhury, M.}, \bibinfo{author}{Sundaram, H.},
      \bibinfo{author}{John, A.} \& \bibinfo{author}{Seligmann, D.~D.}
      \newblock \bibinfo{title}{Can blog communication dynamics be correlated with
      stock market activity?}
      \newblock In \emph{\bibinfo{booktitle}{Proceedings of the nineteenth ACM
      conference on Hypertext and hypermedia}}, \bibinfo{pages}{55--60}
      (\bibinfo{organization}{ACM}, \bibinfo{year}{2008}).

    \bibitem{mao2012correlating}
      \bibinfo{author}{Mao, Y.}, \bibinfo{author}{Wei, W.}, \bibinfo{author}{Wang,
      B.} \& \bibinfo{author}{Liu, B.}
      \newblock \bibinfo{title}{{Correlating S\&P 500 stocks with Twitter data}}.
      \newblock In \emph{\bibinfo{booktitle}{Proceedings of the First ACM
	International Workshop on Hot Topics on Interdisciplinary Social Networks
      Research}}, \bibinfo{pages}{69--72} (\bibinfo{organization}{ACM},
      \bibinfo{year}{2012}).

    \bibitem{ruiz2012correlating}
      \bibinfo{author}{Ruiz, E.~J.}, \bibinfo{author}{Hristidis, V.},
      \bibinfo{author}{Castillo, C.}, \bibinfo{author}{Gionis, A.} \&
      \bibinfo{author}{Jaimes, A.}
      \newblock \bibinfo{title}{Correlating financial time series with micro-blogging
      activity}.
      \newblock In \emph{\bibinfo{booktitle}{Proceedings of the fifth ACM
      international conference on Web search and data mining}},
      \bibinfo{pages}{513--522} (\bibinfo{organization}{ACM},
      \bibinfo{year}{2012}).

    \bibitem{bollen2011twitter}
      \bibinfo{author}{Bollen, J.}, \bibinfo{author}{Mao, H.} \&
      \bibinfo{author}{Zeng, X.}
      \newblock \bibinfo{title}{{Twitter mood predicts the stock market}}.
      \newblock \emph{\bibinfo{journal}{Journal of Computational Science}}
      \textbf{\bibinfo{volume}{2}}, \bibinfo{pages}{1--8} (\bibinfo{year}{2011}).

    \bibitem{bollen2011modeling}
      \bibinfo{author}{Bollen, J.}, \bibinfo{author}{Pepe, A.} \&
      \bibinfo{author}{Mao, H.}
      \newblock \bibinfo{title}{{Modeling public mood and emotion: Twitter sentiment
      and socio-economic phenomena}}.
      \newblock In \emph{\bibinfo{booktitle}{Proceedings of the Fifth International
      AAAI Conference on Weblogs and Social Media}}, \bibinfo{pages}{450--453}
      (\bibinfo{year}{2011}).

    \bibitem{zhang2011predicting}
      \bibinfo{author}{Zhang, X.}, \bibinfo{author}{Fuehres, H.} \&
      \bibinfo{author}{Gloor, P.~A.}
      \newblock \bibinfo{title}{Predicting stock market indicators through {T}witter
    “i hope it is not as bad as i fear”}.
    \newblock \emph{\bibinfo{journal}{Procedia-Social and Behavioral Sciences}}
    \textbf{\bibinfo{volume}{26}}, \bibinfo{pages}{55--62}
    (\bibinfo{year}{2011}).

  \bibitem{zheludev2014can}
    \bibinfo{author}{Zheludev, I.},
    \bibinfo{author}{Smith, R.} \&
    \bibinfo{author}{Aste, T.}
    \newblock
    \bibinfo{title}{When can social media lead financial markets?}.
    \newblock
    \emph{\bibinfo{journal}{Sci. Rep.}}
    \textbf{\bibinfo{volume}{4}} (\bibinfo{year}{2014}).

  \bibitem{preis2012quantifying}
    \bibinfo{author}{Preis, T.}, \bibinfo{author}{Moat, H.~S.},
    \bibinfo{author}{Stanley, H.~E.} \& \bibinfo{author}{Bishop, S.~R.}
    \newblock \bibinfo{title}{Quantifying the advantage of looking forward}.
    \newblock \emph{\bibinfo{journal}{Scientific Reports}}
    \textbf{\bibinfo{volume}{2}} (\bibinfo{year}{2012}).

  \bibitem{preis2013quantifying}
    \bibinfo{author}{Preis, T.}, \bibinfo{author}{Moat, H.~S.} \&
    \bibinfo{author}{Stanley, H.~E.}
    \newblock \bibinfo{title}{Quantifying trading behavior in financial markets
      using {G}oogle {T}rends}.
      \newblock \emph{\bibinfo{journal}{Scientific Reports}}
      \textbf{\bibinfo{volume}{3}} (\bibinfo{year}{2013}).

    \bibitem{moat2013quantifying}
      \bibinfo{author}{Moat, H.~S.} \emph{et~al.}
      \newblock \bibinfo{title}{Quantifying wikipedia usage patterns before stock
      market moves}.
      \newblock \emph{\bibinfo{journal}{Scientific Reports}}
      \textbf{\bibinfo{volume}{3}} (\bibinfo{year}{2013}).

    \bibitem{granger69}
      \bibinfo{author}{Granger, C.}
      \newblock \bibinfo{title}{Investigating causal relations by econometric models
      and cross-spectral methods}.
      \newblock \emph{\bibinfo{journal}{Econometrica}} \textbf{\bibinfo{volume}{37}},
      \bibinfo{pages}{424--438} (\bibinfo{year}{1969}).

    \bibitem{zipf49}
      \bibinfo{author}{Zipf, G.}
      \newblock \emph{\bibinfo{title}{Human Behavior and the Principle of Least
      Effort}} (\bibinfo{publisher}{Addison-Wesley}, \bibinfo{address}{Cambridge,
      MA}, \bibinfo{year}{1949}).

    \bibitem{bormetti2013modelling}
      \bibinfo{author}{Bormetti, G.} \emph{et~al.}
      \newblock \bibinfo{title}{Modelling systemic price cojumps with {H}awkes factor
    models}.
    \newblock \emph{\bibinfo{journal}{Quantitative Finance}}
    \textbf{\bibinfo{volume}{15}}, \bibinfo{pages}{1137--1156} (\bibinfo{year}{2015}).

  \bibitem{calcagnile2015}
    \bibinfo{author}{Calcagnile, L.~M.} \emph{et~al.}
    \newblock \bibinfo{title}{Collective synchronization and high frequency systemic instabilities in financial markets}.
    \newblock \emph{\bibinfo{journal}{arXiv preprint arXiv:1505.00704}}
    (\bibinfo{year}{2015}).

  \bibitem{Clauset_etal:2009}
    \bibinfo{author}{Clauset, A.}, \bibinfo{author}{Shalizi, C.~R.} \&
    \bibinfo{author}{Newman, M.~E.}
    \newblock \bibinfo{title}{Power-law distributions in empirical data}.
    \newblock \emph{\bibinfo{journal}{SIAM review}} \textbf{\bibinfo{volume}{51}},
    \bibinfo{pages}{661--703} (\bibinfo{year}{2009}).

\end{thebibliography}
\end{document}